\documentclass[aps,pre,twocolumn,superscriptaddress,floatfix,nobibnotes,a4paper]{revtex4-2}
\usepackage{graphicx}
\graphicspath{{fig/}}
\usepackage{bm}
\usepackage{amsmath}
\usepackage{url}

\begin{document}

\title{Dynamical scaling study for the estimation of dynamical exponent $z$ of three-dimensional XY spin glass model}

\author{Yusuke Terasawa}
\author{Yukiyasu Ozeki}
\affiliation{Department of Engineering Science, Graduate School of Informatics and Engineering, The University of Electro-Communications, 1-5-1 Chofugaoka, Chofu-shi, Tokyo 182-8585, Japan}

\begin{abstract}
To analyze the $\pm J$ XY spin-glass in three dimensions, we verified a method aimed at obtaining a high-precision dynamical exponent $z$ from the correlation length in the nonequilibrium relaxation process. The obtained $z$ yielded consistent and highly accurate results in previous studies for relatively well-studied models---specifically, the three-dimensional (3D) ferromagnetic Ising model and the 3D $\pm J$ Ising model. Building on these previous studies, we used this method and the dynamical scaling method to analyze the 3D $\pm J$ XY model and obtained highly precise critical temperatures and exponents. The estimated transition temperatures satisfy $T_{\mathrm{SG}}<T_{\mathrm{CG}}$ and are consistent with the spin chirality decoupling picture.
\end{abstract}

\maketitle

\section{Introduction}
Spin-glass (SG) systems have been the subject of extensive study for many years, serving as a prototype for random systems and finding wide-ranging applications in areas such as quantum annealing and information sciences. Despite this extensive research, numerous aspects of the nature of SG phase transitions remain poorly understood. One such aspect pertains to the applicability of a spin chirality decoupling picture in explaining experimental Heisenberg-like SG magnets\cite{Kawamura1992} (i.e., whether $T_{\mathrm{CG}}>T_{\mathrm{SG}}>0$ or $T_{\mathrm{CG}}=T_{\mathrm{SG}}>0$).

The lack of progress in SG research is attributable to the substantial computational time required to achieve equilibrium states as a result of the slow dynamics arising from frustration in SG systems. One effective approach to overcoming this issue is the nonequilibrium relaxation (NER) method. A recent study using this method confirmed the SG universality in the $\pm J$ Ising model in three dimensions\cite{Terasawa2023}.
In the NER method, obtaining the SG universality expressed by critical exponents $\nu$ necessitates the precise determination of the dynamical critical exponent $z$. The authors of a previous study used the asymptotic behavior of the dynamical Binder parameter $g_{\mathrm{SG}}$ (expressed as $g_{\mathrm{SG}}\sim t^{d/z}$, where $t$ denotes time and $d$ represents the dimensionality of the system) at the critical temperature\cite{Blundell1992}. By extrapolating the time $t$ to infinity, the authors were able to evaluate $z$ with high precision.
 However, this method is not applicable when the dynamical Binder parameter $g_{\mathrm{SG}}$ does not exhibit algebraic divergence\cite{Yamamoto2004}. One such instance is observed in the Binder parameter for a chiral glass in the 3D $\pm J$ XY model. To confirm universality in such SG systems, it remains imperative to compute $z$ with high precision even in such cases.\par

 Typically, $z$ is defined using the dynamical correlation length $\xi(t)$ at the critical temperature, defined as:
 \begin{equation}
     \label{z definition}
     \xi(t)\sim t^{1/z}.
 \end{equation}
 Thus, if $\xi(t)$ can be evaluated, then $z$ can be estimated. The second moment method is commonly used to obtain the ``static" correlation length, and this approach has yielded significant results in equilibrium Monte Carlo simulations.
 However, Nakamura has pointed out that even applying the second moment method to estimate the ``dynamical" correlation length $\xi(t)$ may not lead to a precise determination of $z$: Even in simple 3D ferromagnetic (FM) Ising models, the results often deviate substantially from reliable benchmark values\cite{Nakamura2010}. This discrepancy might be attributable to the nontrivial question of whether the second moment correlation length can be applied in nonequilibrium relaxation processes\cite{Nakamura2010}.\par

Recently, a method has been devised to estimate $\xi(t)$ using scaling laws for the dynamical correlation function $f(r,t)$ in conjunction with Gaussian process regression\cite{Nakamura2016,Nakamura2019}. Using this approach to obtain $\xi(t)$, in conjunction with Eq. (\ref{z definition}), Nakamura confirmed that $z$ for the Heisenberg SG model can be computed\cite{Nakamura2019}. Hence, this method is expected to enable $z$ to be calculated for various models, including those encompassing SG systems. However, a detailed validation of this assertion remains largely unexplored.\par

This paper has two main objectives. The first objective is to investigate the critical phenomena in the $\pm J$ XY model, which is an SG system in which the Binder parameter exhibits no algebraic divergence\cite{Yamamoto2004}.
This approach enables an analysis of critical phenomena in various SG systems, such as Heisenberg systems and gauge glass models.
The second objective is to establish the reliability of the estimated $z$ through our study because a precise value of $z$ is needed to study the SG systems through dynamical scaling analysis. To establish the reliability of the estimated $z$, we apply this method to relatively well-studied models---the 3D FM Ising model and the 3D $\pm J$ Ising model---and compute $z$.

This paper is structured as follows. In Sec. 2, we describe the models, observables, and methods used to estimate the critical temperature and dynamical exponent. In Sec. 3, we validate the estimation method of the dynamical exponent by applying it to the 3D FM Ising model and the 3D ±J Ising model. In Sec. 4, we apply this method to the 3D ±J XY model and analyze its critical behavior using dynamical scaling. Finally, Sec. 5 provides a summary and discussion.

\section{Methods}
\label{sec:Methods}
\subsection{Models and Observables for Estimating the Critical Temperature}
\label{sec:Methods:A}

In the present study, we mainly aimed to analyze the $\pm J$ XY model in three dimensions. The Hamiltonian can be expressed as
\begin{equation}
    H=-\sum_{\langle i,j\rangle}J_{ij}\boldsymbol{S}_i\cdot\boldsymbol{S}_j=-\sum_{\langle i,j\rangle}J_{ij}\cos(\theta_i-\theta_j).
\end{equation}
The interaction $J_{ij}$ takes one of two values, $+J$ or $-J$, with the same probability. The summation $\sum_{\langle i,j\rangle}$ is taken over all nearest-neighboring sites. The temperature $T$ is scaled by $J/k_{\mathrm{B}}$. Each spin $\boldsymbol{S}_i$ is characterized by an angle $\theta_i$ in the $XY$ plane. The linear lattice size is denoted by $L$. The total number of spins is $N=L^2(L-1)$, and a skewed periodic boundary condition is imposed. \par

 We calculated the SG susceptibility $\chi_{\rm SG}$ for SG systems. It is defined as
 \begin{equation}
     \chi_{\rm SG}=\frac{1}{N}\Bigg[\sum_{i,j}\langle \boldsymbol{S}_i\cdot\boldsymbol{S}_j\rangle^2\Bigg]_J,
 \end{equation}
where $\langle\cdots\rangle$ denotes the thermal average and $[\cdots]_J$ denotes the average for random-bond configurations. To calculate this quantity, the thermal average is replaced by the average over independent real replicas that consist of different thermal ensembles:
\begin{equation}
    \langle \boldsymbol{S}_i\boldsymbol{S}_j\rangle=\frac{1}{n}\sum_{A=1}^{n}\boldsymbol{S}_i^{(A)}\boldsymbol{S}_j^{(A)},
\end{equation}
where $n$ denotes the replica number and $A$ denotes the replica index. The magnitude of $n$ affects the accuracy of thermal averages. Each replica system has the same bonds $J_{ij}$ but starts with a different randomly assigned initial spin state.
The replicas' overlap between $A$ and $B$ is defined by
\begin{equation}
    \label{eq: overlap}
    q_{\mu\nu}^{AB}=\frac{1}{N}\sum_i^{N} S_{i\mu}^{(A)}S_{i\nu}^{(B)}.
\end{equation}
where subscripts $\mu$ and $\nu$ denote the components of spin. The SG susceptibility can be rewritten using this overlap as
\begin{equation}
    \label{chiSG produced by replica overlap}
    \chi_{\mathrm{SG}}=\frac{N}{C_n}\Bigg[\sum_{A>B}\sum_{\mu,\nu}(q_{\mu\nu}^{AB})^2\Bigg]_J.
\end{equation}
Here, $C_n=n(n-1)/2$ represents the number of combinations for choosing two replicas out of $n$ ones. \par
The existence of the chiral-glass (CG) transition in the 3D $\pm J$ XY model has been confirmed in numerous previous studies\cite{Kawamura2001,Yamamoto2004,Nakamura2006}.
The CG susceptibility $\chi_{\mathrm{CG}}$ is similarly expressed as
\begin{equation}
    \label{chiCG produced by replica overlap}
    \chi_{\mathrm{CG}}=\frac{3N}{C_n}\Bigg[\sum_{A>B}(q_\kappa^{AB})^2\Bigg]_J,
\end{equation}
where
\begin{equation}
    q_\kappa^{AB}=\frac{1}{3N}\sum_{\alpha}^{3N}\kappa_{\alpha}^{(A)}\kappa_{\alpha}^{(B)},
\end{equation}
\begin{equation}
    \kappa_{\alpha}^{(A)}=\frac{1}{2\sqrt{2}}\sum_{\langle jk\rangle\in\alpha}J_{jk}\sin(\theta_j^{(A)}-\theta_k^{(A)}).
\end{equation}
Here, at each square plaquette $\alpha$ composed of nearest-neighbor bonds $\langle jk\rangle$, a local chirality variable $\kappa_{\alpha}$ is defined.\par

To estimate the critical temperature and some critical exponents of the 3D $\pm J$ XY model, we used the following dynamical scaling law for SG susceptibility\cite{Yamamoto2004,Nakamura2006,Terasawa2023}:
\begin{equation}
    \label{dynamical scaling law for critical temperature SG}
    \chi_{\mathrm{SG}}(t,T)=t^{\lambda_{\mathrm{SG}}(T)}\Phi\bigg(\frac{t}{\tau_{\mathrm{SG}}(T)}\bigg),
\end{equation}
where
 \begin{equation}
   \lambda_{\mathrm{SG}}(T)=\frac{\gamma}{z_{\mathrm{s,SG}}(T)\nu},
 \end{equation}
 \begin{equation}
   \label{tau scaling SG}
   \tau_{\mathrm{SG}}(T)=(T-T_{\mathrm{SG}})^{-z_{\mathrm{s,SG}}(T)\nu}.
 \end{equation}
\begin{equation}
    \label{z assumption SG}
    z_{\mathrm{s,SG}}(T)=z\frac{T_{\mathrm{SG}}}{T}.
\end{equation}

And for CG case, we used the following dynamical scaling law
\begin{equation}
    \label{dynamical scaling law for critical temperature CG}
    \chi_{\mathrm{CG}}(t,T)=t^{\lambda_{\mathrm{CG}}(T)}\Phi\bigg(\frac{t}{\tau_{\mathrm{CG}}(T)}\bigg),
\end{equation}
where
 \begin{equation}
   \lambda_{\mathrm{CG}}(T)=\frac{\gamma}{z_{\mathrm{s,CG}}(T)\nu},
 \end{equation}
 \begin{equation}
   \label{tau scaling CG}
   \tau_{\mathrm{CG}}(T)=(T-T_{\rm CG})^{-z_{\mathrm{s,CG}}(T)\nu}.
 \end{equation}
\begin{equation}
    \label{z assumption CG}
    z_{\mathrm{s,CG}}(T)=z\frac{T_{\mathrm{CG}}}{T}.
\end{equation}

The relation expressed by Eq. (\ref{z assumption SG}) holds only within the context of the NER process in SG systems. Just above the transition temperature, Eq. (\ref{z assumption SG}) becomes equivalent to the definition of $z$. This relation was first noted for the Gaussian bond Ising model and gauge glass model\cite{Katzgraber2005}. It was later noted for Ising, XY, and Heisenberg SG systems\cite{Nakamura2006}. The crossover between this dynamics and the normal dynamics is observed at the multicritical point in Ising SG systems\cite{Terasawa2023}.
However, while the temperature dependence of $z(T)$ has been widely discussed and generally supported, there remain some differences in the detailed behavior reported in the literature, particularly regarding its quantitative form over the entire temperature range\cite{Nakamura2006,Nakamura2019}.
An analysis based on the aforementioned method has been previously reported\cite{Nakamura2006}. Considering the advances in numerical techniques and available computational power since then, we conducted renewed simulations in the present study.

\subsection{Observables and Scaling Law for Estimating the Dynamical Critical Exponent}
In the scaling analysis based on the method described in Sec. \ref{sec:Methods:A}, the critical temperature $T_{\mathrm{c}}$ and the products of the critical exponents $\gamma/z\nu$ and $z\nu$ can be obtained. Therefore, to estimate all of the critical exponents, it is necessary to calculate $z$. On the basis of Eq. (\ref{z definition}), we will explain the method for determining the dynamical correlation length $\xi(t)$ required to calculate $z$.

The SG correlation function $f_{\mathrm{SG}}(r)$ is defined by
\begin{equation}
    f_{\mathrm{SG}}(r)=\Bigg[\frac{1}{N}\sum_i^{N}\langle \boldsymbol{S}_i\cdot\boldsymbol{S}_{i+r}\rangle^2\Bigg]_J,
\end{equation}
where $r$ denotes the distance between two spins in the lattice. This equation can be rewritten using Eq. (\ref{eq: overlap}) as
\begin{equation}
    \label{deformed fSG 1}
    f_{\mathrm{SG}}(r) = \frac{1}{NC_n}\Bigg[\sum_{A>B}\sum_{i,\mu,\nu}(q_{\mu\nu}^{AB}(i))(q_{\mu\nu}^{AB}(i+r))\Bigg]_J ,
\end{equation}
which is practically expressed as
\begin{equation}
    \label{deformed fSG 2}
    \begin{aligned}[t]
        f_{\mathrm{SG}}(r)=\frac{1}{N}\Bigg[ \sum_{i}^{N}&\Bigg\{\Bigg(\frac{1}{n}\sum_{A=1}^{n}\boldsymbol{S}_i^{(A)}\cdot\boldsymbol{S}_{i+r}^{(A)}\Bigg)^2 \\
        &-\frac{1}{n^2}\sum_{A=1}^{n}\Bigg(\boldsymbol{S}_i^{(A)}\cdot\boldsymbol{S}_{i+r}^{(A)}\Bigg)^2 \Bigg\}
        \Bigg]_J ,
    \end{aligned}
\end{equation}
for computations.
This choice was made because of the computational complexity associated with Eq. (\ref{deformed fSG 1}), which requires computations in $_nC_2$ different ways.\par

Similarly, the CG correlation function $f_{\mathrm{CG}}$ is also defined by
\begin{equation}
    \begin{aligned}[b]
        f_{\mathrm{CG}} &= \frac{1}{3N}\Bigg[ \sum_{\alpha}^{3N}\Bigg\{\Bigg(\frac{1}{n}\sum_{A=1}^{n}\kappa_\alpha^{(A)}\kappa_{\alpha+r}^{(A)}\Bigg)^2 \\
        &\qquad -\frac{1}{n^2}\sum_{A=1}^{n}\Bigg(\kappa_\alpha^{(A)}\kappa_{\alpha+r}^{(A)}\Bigg)^2 \Bigg\}
    \Bigg]_J.
    \end{aligned}
\end{equation}
For both SG and CG correlation functions, we assume the dynamical scaling ansatz\cite{Nakamura2016,Nakamura2019} expressed as

\begin{equation}
    \label{correlation dynamical scaling law}
    \frac{f(r,t)}{ \xi(t)^{-d+2-\eta_{\mathrm{eff}}} }=\Phi\bigg(\frac{r}{\xi(t)}\bigg).
\end{equation}
Here, $d$ is the spatial dimension, $\eta_{\mathrm{eff}}$ is an effective scaling exponent, and $f(r,t)$ and $\xi(t)$ are the correlation function and correlation length, respectively, at Monte Carlo step (MCS) $t$. We estimate $\eta_{\mathrm{eff}}$ and dozens of $\xi(t)$ such that all of the $f(r,t)$ data collapse onto a single scaling function $\Phi(\cdot)$.

The concrete procedure for estimating $\xi(t)$ and $z$ in this method is as follows:
\begin{enumerate}
\item At the transition temperature, we compute $f(r,t)$ and obtain the initial estimates of $\{\xi(t)\}$ using the second-moment method.
\item Using the obtained $f(r,t)$, we perform an optimization in which the initial values of $\{\xi(t)\}$ are given by the second-moment estimates, and determine the parameters that satisfy the dynamical scaling ansatz Eq.~(\ref{correlation dynamical scaling law}).
\item From the optimized $\{\xi(t)\}$, we construct a log--log plot of $\xi(t)$ versus the MCS step $t$, and estimate $z$ from its slope together with Eq.~(\ref{z definition}).
\end{enumerate}

The second-moment correlation lengths used in the above procedure are given by
\begin{equation}
    \label{eq: SG second moment correlation length}
    \xi_{\mathrm{2nd,SG}}=\frac{1}{2\sin{(\pi/L)}}\sqrt{\frac{\tilde{\chi}_{\mathrm{SG}}(\mathbf{0})}{\tilde{\chi}_{\mathrm{SG}}(\mathbf{k}_{\mathrm{min}})}-1},
\end{equation}
\begin{equation}
    \label{eq: CG second moment correlation length}
    \xi_{\mathrm{2nd,CG}}=\frac{1}{2\sin{(\pi/L)}}\sqrt{\frac{\tilde{\chi}_{\mathrm{CG}}(\mathbf{0})}{\tilde{\chi}_{\mathrm{CG}}(\mathbf{k}_{\mathrm{min}})}-1},
\end{equation}
where $\tilde{\chi}_{\mathrm{SG}}(\cdot)$ and $\tilde{\chi}_{\mathrm{CG}}(\cdot)$ represent the Fourier transform of susceptibilities and $\mathbf{k}_{\mathrm{min}}$ denotes the smallest wave vector, which is determined as $\mathbf{k}_{\mathrm{min}}=(2\pi/L,0,0)$ in the present study.\par

The second moment correlation length can be calculated during the nonequilibrium relaxation process. However, the $z$ estimated using this method yields inaccurate values even for relatively simple models such as the 3D FM Ising model\cite{Nakamura2010}. Therefore, an analysis based on Eq. (\ref{correlation dynamical scaling law}) is necessary.

\subsection{Gaussian Process Regression}
In this subsection, we briefly explain the method of dynamic scaling analysis based on Gaussian process regression(GPR)~\cite{Harada2011,Echinaka2016,Nakamura2016,Nakamura2019}. This method is applied in all scaling analyses, and it is also used in Sec.~\ref{subsection: Critical Temperatures from Dynamical Scaling Analysis} to assess the validity of the scaling plots. Let us rewrite the scaling relations expressed in Eqs. (\ref{dynamical scaling law for critical temperature SG}), (\ref{dynamical scaling law for critical temperature CG}), and (\ref{correlation dynamical scaling law}) in the following form:
\begin{equation}
    Y = \Phi(X).
\end{equation}
We denote the data vectors formed by $X$ and $Y$ as $\vec{X}$ and $\vec{Y}$, respectively, and the dynamic scaling function evaluated at $\vec{X}$ as $\vec{\Phi}$. Then, the log-likelihood function for optimizable parameters $\vec{\theta}$ is given by
\begin{equation}
    \label{GPR Loglikelihood}
    \ln{\mathcal{L}(\vec{\theta})} = -\frac{1}{2}\ln{|2\pi\Sigma|} - \frac{1}{2}(\vec{Y} - \vec{\Phi})^t \Sigma^{-1} (\vec{Y} - \vec{\Phi}),
\end{equation}
where $\Sigma$ is the covariance matrix whose elements are defined via a kernel function $K(X_i, X_j, \vec{\theta})$ as
\begin{equation}
    \Sigma_{ij} = E_i^2 \delta_{ij} + K(X_i, X_j, \vec{\theta}),
\end{equation}
\begin{equation}
    K(X_i, X_j, \vec{\theta}) = \theta_0^2 \delta_{ij} + \theta_1^2 \exp\left(-\frac{(X_i - X_j)^2}{2\theta_2^2} \right).
\end{equation}
Here, $E_i$ represents the error of $Y_i$, and $(\theta_0, \theta_1, \theta_2)$ are the hyperparameters. By maximizing Eq.~(\ref{GPR Loglikelihood}) using an optimization algorithm, the parameters can be estimated. Specifically, the dynamic scaling analysis using Eqs.~(\ref{dynamical scaling law for critical temperature SG}) and (\ref{dynamical scaling law for critical temperature CG}) yield the transition temperature and the exponents $\gamma / z\nu$ and $z\nu$, while the analysis using Eq.~(\ref{correlation dynamical scaling law}) provides the sets of $\xi(t)$ and $\eta_{\mathrm{eff}}$.

In the actual analysis, we applied a logarithmic transformation to the data to stabilize the fitting and improve numerical accuracy\cite{Echinaka2016}, defined as follows:
\begin{align}
    X_i' &\equiv \ln X_i , \notag \\
    Y_i' &\equiv \ln Y_i , \notag \\
    E_i' &\equiv \ln \left( 1 + \frac{E_i}{Y_i} \right) \approx \frac{E_i}{Y_i} .
\end{align}

\section{Confirmation for the Estimation of the Dynamical Exponent}
\label{section: estimation of z}
To calculate $z$, we first perform a scaling analysis based on Eq. (\ref{correlation dynamical scaling law}) to obtain a set of dynamical correlation lengths $\xi(t)$ at the transition temperature and then estimate $z$ from the slope of $\xi(t)$ versus $t$ on a double-logarithmic plot according to Eq. (\ref{z definition}). However, the reliability of obtained $z$ values using this method has not been thoroughly validated. Therefore, we first apply this method to calculate $z$ for the relatively well-studied 3D FM Ising model and the $\pm J$ Ising model to confirm if the results are consistent with previous studies and establish the reliability of this method. Following this confirmation, we apply the method to the primary focus of the present study, the 3D $\pm J$ XY model in the next section.

\subsection{3D FM Ising Model}\label{demonstration 3D ferro Isingmodel}
\begin{figure}[tbp]
    \centering
    \includegraphics[width=0.98\linewidth]{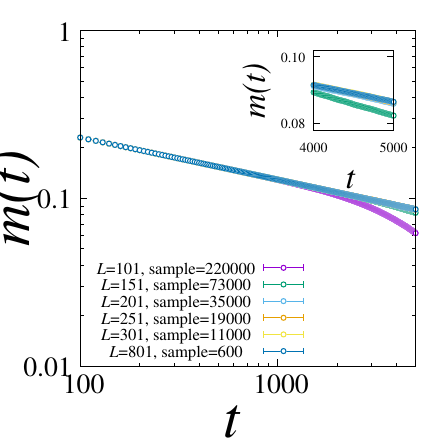}
    \caption{Size dependence of the magnetization in the 3D ferromagnetic Ising model.
The simulations were performed at the transition temperature $T_{\mathrm{c}} = 4.51152325$\,\cite{Ferrenberg2018}.
The number of samples is indicated in the legend.
It is found that the size dependence can be neglected for $L \ge 301$.}

    \label{fig:FM Ising Size dependence}
\end{figure}

To analyze the 3D FM Ising model by the dynamical scaling method, we conducted simulations up to 5000 MCS.
To examine the size dependence in this range, we calculated the relaxation process of the magnetization from the ordered state near the transition temperature.
When the size dependence is negligible, the magnetization exhibits a power-law decay behavior\cite{Ozeki_2007}.
The transition temperature was chosen as the highly accurate value obtained from a previous study, $T_{\mathrm{c}} = 4.51152325$\cite{Ferrenberg2018}.
The results are shown in Fig.~\ref{fig:FM Ising Size dependence}, indicating that the size dependence can be neglected for $L \ge 301$ within this MCS range.

We therefore adopted a linear size of $L=801$ in the present study. Under this condition, we computed the correlation function $f(r,t)$ at the critical temperature. Here, the sample number was set to 1,200.
In addition, the MCSs for computing the correlation function were selected to be logarithmically spaced with 10 points between 100 MCS and 5000 MCS.\par

\begin{figure}[tbp]
    \centering
    \includegraphics[width=0.98\linewidth]{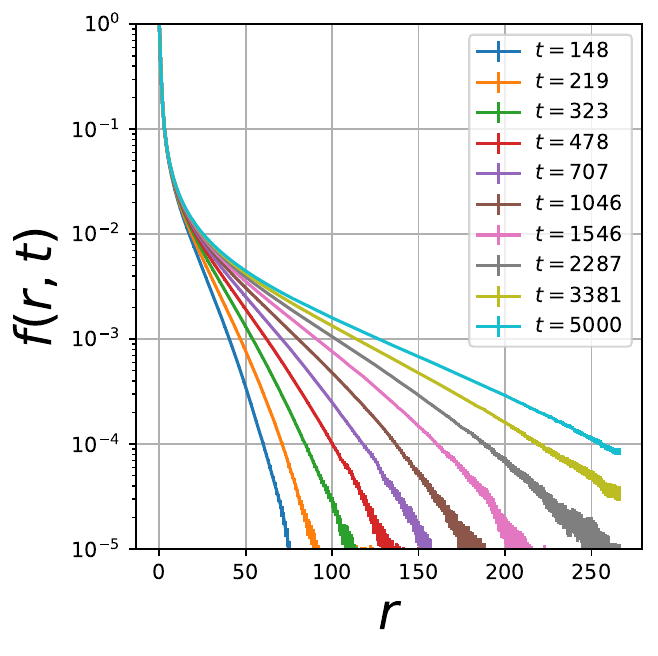}
    \caption{The dynamical correlation function of the 3D FM Ising model, $f(r,t)$, at the critical temperature $T_{\mathrm{C}}=4.51152325$\cite{Ferrenberg2018} for various Monte Carlo steps $t$.}
    \label{fig:3dising_corfunction}
\end{figure}

\begin{figure}[tbp]
    \centering
    \includegraphics[width=0.98\linewidth]{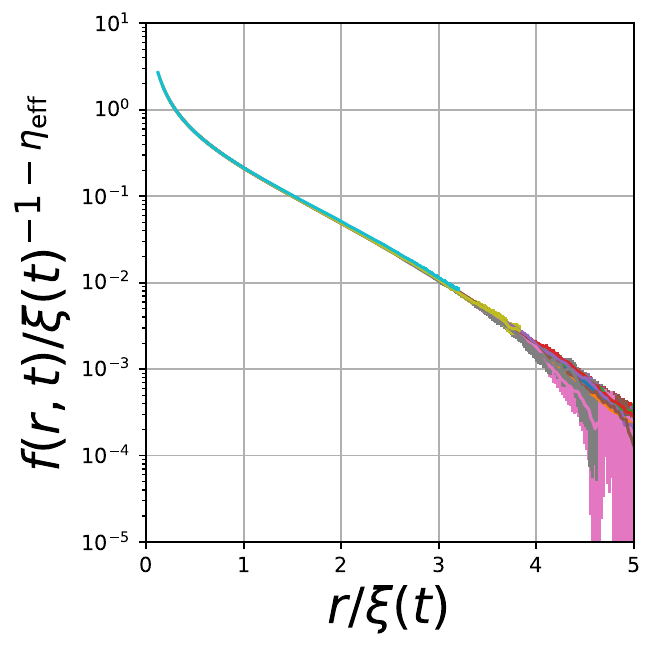}
    \caption{The scaling plot of the dynamical correlation function $f(r,t)$ for the 3D FM Ising model, based on the dynamical scaling law in Eq. (\ref{correlation dynamical scaling law}). The quantities used in the scaling plot are the effective exponent $\eta_{\mathrm{eff}}=0.036(1)$ and the dozens of correlation length $\{\xi(t)\}$ shown in Fig. \ref{fig:3dising_determinateZ}. These values were obtained through optimization calculations of Gaussian process regression.}
    \label{fig:3dising_scaling}
\end{figure}

Figure \ref{fig:3dising_corfunction} shows the dynamical correlation function of the 3D FM Ising model at the critical temperature. We subsequently conducted a dynamical scaling analysis based on Eq. (\ref{correlation dynamical scaling law}) to determine the set of correlation lengths required for an accurate estimation of $z$. In an optimization procedure, we discarded short-range correlations ($r<10$) and data of small $f(r,t)$ values ($f(r,t)<10^{-4}$). The initial values of $\{\xi(t)\}$ were set to those of correlations obtained by the second moment method. Figure \ref{fig:3dising_scaling} shows the scaling result based on the dynamical scaling law in Eq. (\ref{correlation dynamical scaling law}).

\begin{figure}[tbp]
    \centering
    \includegraphics[width=0.98\linewidth]{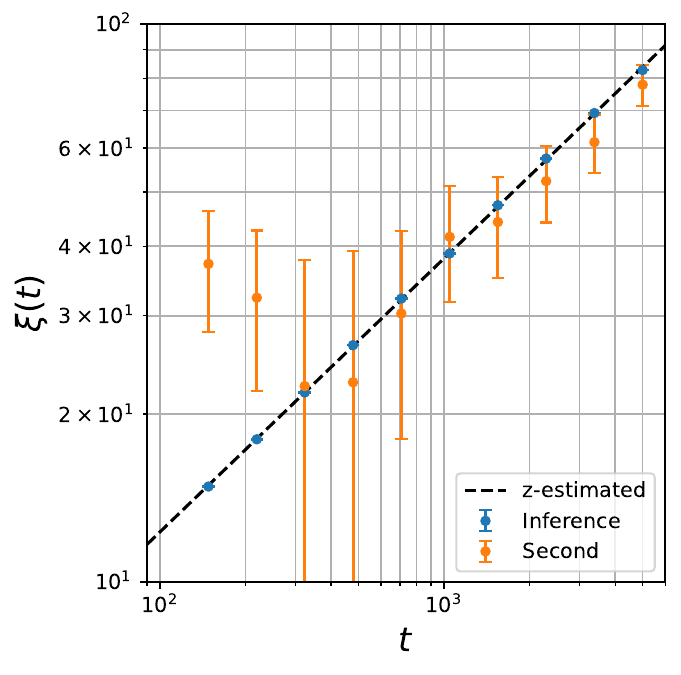}
    \caption{The correlation length $\{\xi(t)\}$ obtained at the critical point of the 3D FM Ising model. Here, ``Inference" in the legend of the graph corresponds to $\{\xi(t)\}$ obtained through optimization calculations of the dynamical scaling law, whereas ``Second" represents $\{\xi(t)\}$ obtained using the second moment method. The dotted line indicates the slope corresponding to the estimated dynamic exponent $z$.
    }
    \label{fig:3dising_determinateZ}
\end{figure}

The resultant values of the correlation length are plotted in Fig. \ref{fig:3dising_determinateZ} together with those obtained by the second-moment method for comparison. From the inference results and Eq.~(\ref{z definition}), we evaluated the dynamic exponent as $z = 2.038(2)$, and the corresponding effective exponent as $\eta_{\mathrm{eff}} = 0.036(1)$. In addition, the uncertainty was estimated as the standard deviation of the bootstrap distribution obtained from 100 resamplings~\cite{Efron1979}.

This value of $z$ is compatible with the high-precision estimate $z = 2.0346(3)$ obtained using the latest reliable methods~\cite{Osada2024}, demonstrating the robustness and accuracy of our scaling analysis based on Eq.~(\ref{correlation dynamical scaling law}).
The obtained value of $\eta_{\mathrm{eff}}$ is also compatible, within error bars, with those reported in reliable previous studies~\cite{Ferrenberg2018}, where $\eta = 0.0361(6)$. Although $\eta_{\mathrm{eff}}$ is introduced as an effective exponent in the present scaling framework, the obtained value of $\eta_{\mathrm{eff}}$ shows good agreement with the previous estimate.

\subsection{3D $\pm J$ Ising Model}\label{demonstration 3D pmj Isingmodel}
Demonstrations of the calculation of $z$ for the 3D $\pm J$ Ising model were performed similarly as described in Sec. \ref{demonstration 3D ferro Isingmodel}.
Simulations were conducted up to $5\times10^5$ MCS. Within this range, finite-size effects were confirmed to vanish for $L\geq31$\cite{Terasawa2023}. In the present study, a linear size of $L=61$ was set, $n=256$ replicas were employed, and the sample number was set to 10. The critical temperature was chosen to be the highly accurate value obtained from a previous study, $T_{\mathrm{SG}}=1.178$\cite{Terasawa2023}. The MC steps for computing the correlation function were selected to be logarithmically spaced with 50 points between 100 MCS and $5\times10^5$ MCS.\par

\begin{figure}[tbp]
    \centering
    \includegraphics[width=0.98\linewidth]{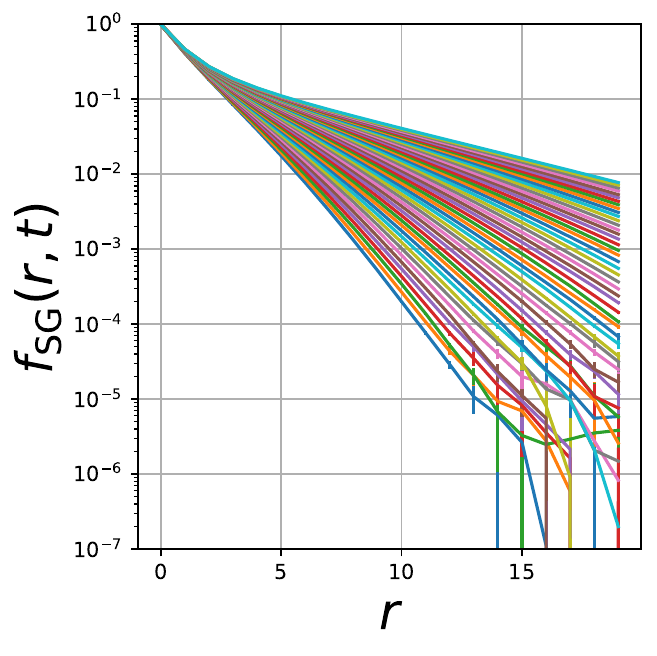}
    \caption{The dynamical SG correlation function of the 3D $\pm J$ Ising model, $f_{\mathrm{SG}}(r,t)$, at the critical temperature $T_{\mathrm{SG}}=1.178$\cite{Terasawa2023} for various Monte Carlo steps $t$.}
    \label{fig:3disingsg_corfunction}
\end{figure}

\begin{figure}[tbp]
    \centering
    \includegraphics[width=0.98\linewidth]{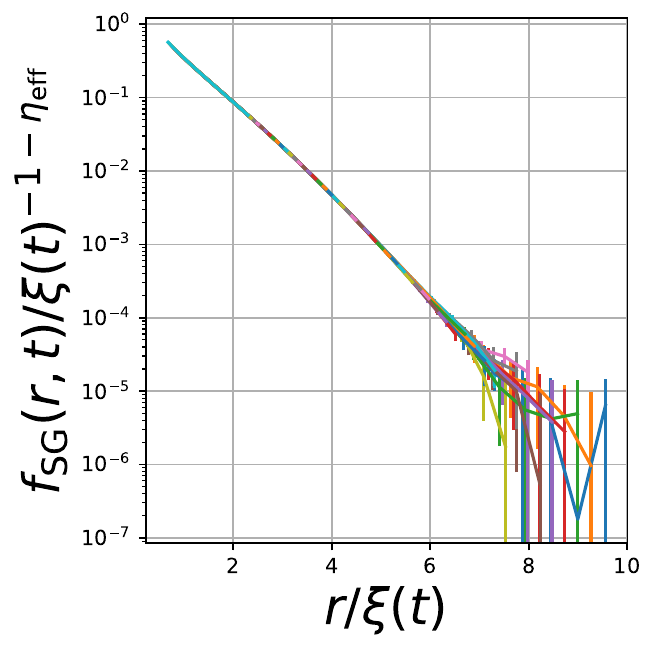}
    \caption{The scaling plot of the dynamical SG correlation function $f_{\mathrm{SG}}(r,t)$ for the 3D $\pm J$ Ising model, based on the dynamical scaling law in Eq. (\ref{correlation dynamical scaling law}). The quantities used in the scaling plot are the effective exponent $\eta_{\mathrm{eff}}=-0.183(2)$ and the dozens of correlation length $\{\xi(t)\}$ shown in Fig. \ref{fig:3disingsg_determinateZ}. These values were obtained through optimization calculations of Gaussian process regression.}
    \label{fig:3disingsg_scaling}
\end{figure}

Figure \ref{fig:3disingsg_corfunction} shows the dynamical SG correlation function of the 3D $\pm J$ Ising model at the critical temperature. We subsequently conducted a dynamical scaling analysis based on Eq. (\ref{correlation dynamical scaling law}) to determine the set of correlation lengths required to accurately estimate $z$. During the optimization procedure, short-range correlations ($r<5$) and data with small $f(r,t)$ values ($f(r,t)<5\times10^{-5}$) were excluded. The initial values of $\{\xi(t)\}$ were set to those of correlations obtained by the second moment method (Eq. (\ref{eq: SG second moment correlation length})).
Figure \ref{fig:3disingsg_scaling} shows the scaling result obtained using the dynamical scaling law in Eq. (\ref{correlation dynamical scaling law}).

\begin{figure}[tbp]
    \centering
    \includegraphics[width=0.98\linewidth]{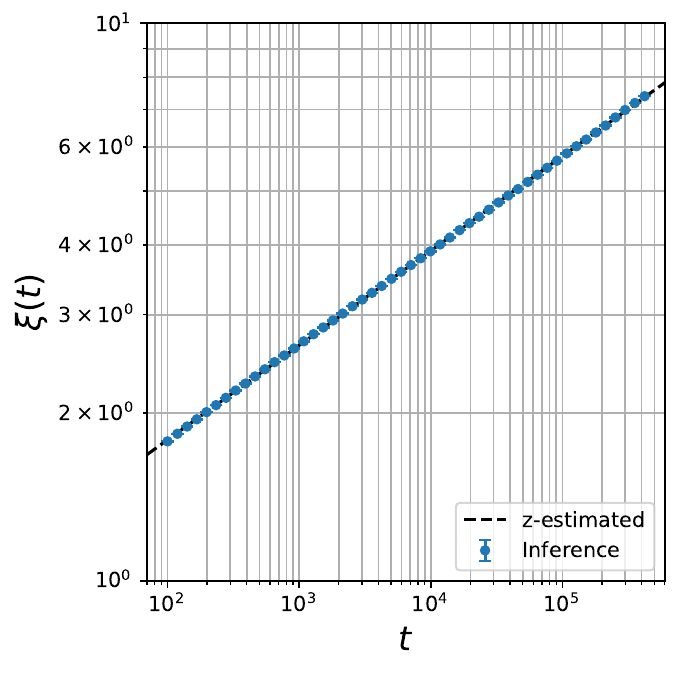}
    \caption{The correlation length $\{\xi(t)\}$, as obtained through optimization calculations of the dynamical scaling law, at the critical point of the $\pm J$ Ising model in three dimensions. The dotted line indicates the slope corresponding to the estimated dynamic exponent $z$. }
    \label{fig:3disingsg_determinateZ}
\end{figure}

The resultant values of correlation length are plotted in Fig. \ref{fig:3disingsg_determinateZ}.
From the inference results and Eq. (\ref{z definition}), we evaluated the value of $z$ to be $z=5.895(7)$; its uncertainty, defined as the standard deviation of the bootstrap distribution, was assessed using 100 bootstrap trials\cite{Efron1979}. This result is consistent with the value $z=5.89(3)$ obtained through extrapolation of the dynamical Binder parameter\cite{Terasawa2023}, demonstrating the robustness and accuracy of our scaling analysis approach based on Eq. (\ref{correlation dynamical scaling law}).
The corresponding effective exponent was evaluated as $\eta_{\mathrm{eff}} = -0.183(2)$. This value is compatible, within error bars, with the previous estimate $\eta = -0.22(4)$ reported in previous study~\cite{Terasawa2023}. Although $\eta_{\mathrm{eff}}$ is introduced as an effective exponent in the present scaling framework, the obtained value of $\eta_{\mathrm{eff}}$ remains consistent with the earlier result.


\section{Results for 3D $\pm J$ XY model}
\subsection{Size-dependence and the behavior of $z(T)$}
\label{subsection:Size-dependence and the behavior of $z(T)$}
To investigate the critical phenomena of SG systems using the dynamical scaling method, it is necessary to perform a scaling analysis based on Eqs. (\ref{dynamical scaling law for critical temperature SG}) and (\ref{dynamical scaling law for critical temperature CG}); this analysis provides the transition temperature and the product of the critical exponents. For this purpose, identifying a linear size $L$ for which finite-size effects can be neglected within the range of considered MCSs is critical.\par

\begin{figure}[tbp]
    \centering
    \begin{minipage}{0.47\linewidth}
        \centering
        \includegraphics[width=\linewidth]{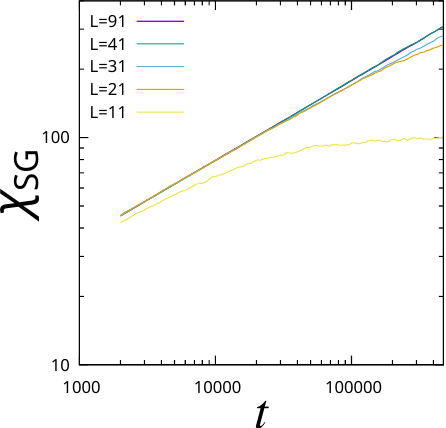}
        \\(a)
    \end{minipage}
    \hfill
    \begin{minipage}{0.47\linewidth}
        \centering
        \includegraphics[width=\linewidth]{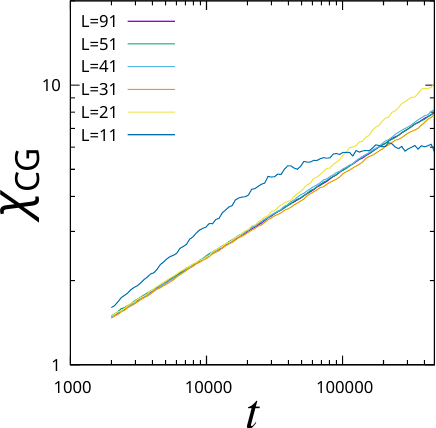}
        \\(b)
    \end{minipage}
    \caption{Size dependence of $\chi_{\mathrm{SG}}$ and $\chi_{\mathrm{CG}}$ in the 3D $\pm J$ XY model: (a) $\chi_{\mathrm{SG}}$, where size dependence vanishes for $L\geq41$, and (b) $\chi_{\mathrm{CG}}$, where size dependence vanishes for $L\geq51$. The temperatures used for the calculations were $T=0.4388$ for $\chi_{\mathrm{SG}}$ and $T=0.4533$ for $\chi_{\mathrm{CG}}$, which are close to the critical point.}
    \label{fig:size dependence of 3D XY SG}
\end{figure}

First, to examine size dependence, we investigated the behavior near the transition temperature of $\chi_{\mathrm{SG}}$ and $\chi_{\mathrm{CG}}$ defined respectively by Eqs. (\ref{chiSG produced by replica overlap}) and (\ref{chiCG produced by replica overlap}). These physical quantities exhibit algebraic divergence at the transition temperature. According to Fig. \ref{fig:size dependence of 3D XY SG}, within the range of $5\times10^{5}$ MCS, the size dependence of $\chi_{\mathrm{SG}}$ disappears for $L\geq41$ and so does that of  $\chi_{\mathrm{CG}}$ for $L\geq51$. On the basis of these results, the size of the 3D $\pm J$ XY model was set to $L=91$ in the present study.\par

Second, to estimate the transition temperature, we have to conduct a scaling analysis using the dynamical scaling law represented by Eqs. (\ref{dynamical scaling law for critical temperature SG}) and (\ref{dynamical scaling law for critical temperature CG}).

To perform this analysis, it is necessary to examine the temperature dependence of $z$ expressed in Eqs.~\ref{z assumption SG} and \ref{z assumption CG}. From the demonstrations described in Secs.~\ref{demonstration 3D ferro Isingmodel} and \ref{demonstration 3D pmj Isingmodel}, we note that the reliability of $z$ obtained from the correlation length estimated by the dynamical scaling law has been established. Based on this result, we calculated the SG and CG correlation functions over a temperature range that includes both low- and high-temperature regimes. Specifically, the temperatures were chosen in the range $0.300 \le T \le 0.600$, using $n = 256$ and 5 samples. The correlation function was computed up to $5 \times 10^5$ MCS.
Subsequently, a scaling analysis was conducted according to Eq.~(\ref{correlation dynamical scaling law}).

In an optimization procedure, we discarded short-range correlations ($r<3$ for $f_{\mathrm{SG}}$ and $r<4$ for $f_{\mathrm{CG}}$) and data of small values ($f_{\mathrm{SG}}(r,t)<10^{-5}$ and $f_{\mathrm{CG}}(r,t)<2\times10^{-6}$).

The dynamical exponent $z$ was then estimated from the slope of $\xi(t)$ versus $t$ on a double-logarithmic plot. The uncertainties were quantified as the standard deviations of the bootstrap distributions derived from 128 resamplings~\cite{Efron1979}.

\begin{figure}[tbp]
    \centering
    \begin{minipage}{0.82\linewidth}
        \centering
        \includegraphics[width=\linewidth]{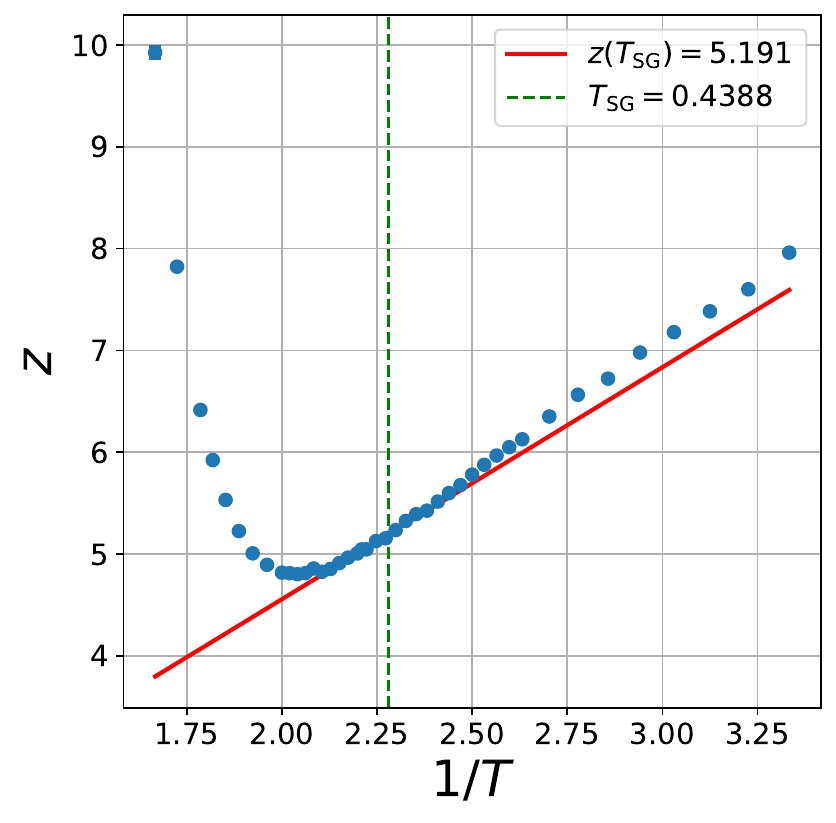}
        \\(a)
    \end{minipage}
    \vspace{2mm}

    \begin{minipage}{0.82\linewidth}
        \centering
        \includegraphics[width=\linewidth]{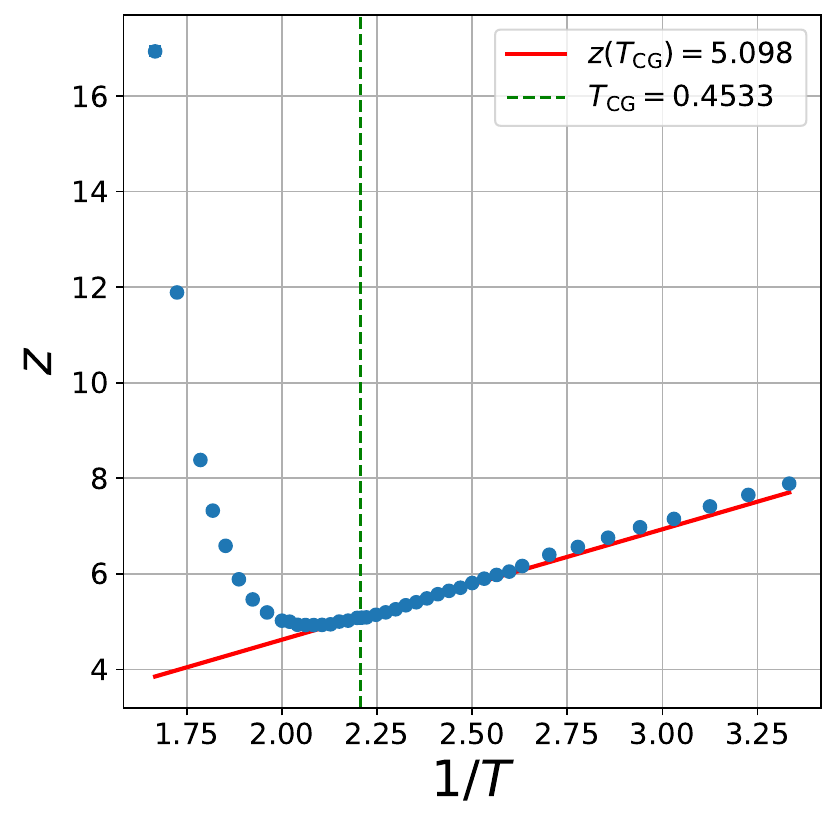}
        \\(b)
    \end{minipage}
    \caption{Temperature dependence of $z$ for the SG and CG transitions. (a) shows the result for the SG case, and (b) shows the result for the CG case. The dashed line indicates the estimated transition temperatures obtained in this study. The linear fits of $z$ are plotted using the transition temperatures and the values of $z$ at the corresponding critical points.}

    \label{1overt dependence}
\end{figure}

The result is shown in Fig.~\ref{1overt dependence}. According to this figure, the temperature dependence of $z$ expressed in Eqs.~\ref{z assumption SG} and \ref{z assumption CG} is confirmed for both SG and CG susceptibilities in the vicinity of the transition temperature. It should be noted that, on the high-temperature side away from the transition temperature, the value of $z$ increases. This behavior arises because, within the time window up to $5 \times 10^5$ MCS, the correlation length exhibits a tendency toward saturation.

We also note that some previous studies have reported different conclusions regarding the temperature dependence of $z$. For example, it has been reported that $z$ becomes constant on the high-temperature side in the case of the CG transition~\cite{Nakamura2006}, and that in the Heisenberg SG system $z$ exhibits a linear temperature dependence with a finite intercept~\cite{Nakamura2019}. Because finite-size effects are sufficiently small in both the previous and present calculations, the difference should not be attributed to the system size. A possible origin is a finite-time correction: within the shorter observation-time range used in Ref.~\cite{Nakamura2006}, the temperature dependence observed in our longer-time data cannot be clearly distinguished. The different representations of the scaling function---an explicit polynomial approximation in Ref.~\cite{Nakamura2006} and GPR in the present work---may also contribute to the difference.

\subsection{Dynamical Scaling Analysis}
\label{subsection: Critical Temperatures from Dynamical Scaling Analysis}

Based on the previously obtained temperature dependence of the dynamical exponent $z(T)$, we performed a dynamical scaling analysis using the relation given in Eqs.~(\ref{dynamical scaling law for critical temperature SG}) and (\ref{dynamical scaling law for critical temperature CG}).

The quantities used in the calculations for both $\chi_{\mathrm{SG}}$ and $\chi_{\mathrm{CG}}$ were $n = 256$ and 5 samples, with simulations performed up to $5 \times 10^{5}$~MCS. The initial temperature range employed in the simulations was $0.300 \le T \le 0.600$.

Next, we examine the temperature range of the data used for the dynamical scaling analysis. The dynamical scaling law expressed by Eqs.~(\ref{dynamical scaling law for critical temperature SG}) and (\ref{dynamical scaling law for critical temperature CG}) may, in principle, hold also on the low-temperature side. However, in the present analysis the scaling fit did not succeed when low-temperature data were included; therefore, in what follows we restrict the analysis to the high-temperature side.

In the SG analysis based on the dynamical scaling law, we take into account the temperature dependence of $z$ assumed in Eq.~(\ref{z assumption SG}) (and similarly for CG in Eq.~(\ref{z assumption CG})). Hence, the temperature range should be chosen so that these assumptions are satisfied. The resulting temperature dependence of $z$ obtained in Sec.~\ref{subsection:Size-dependence and the behavior of $z(T)$} is shown in Fig.~\ref{1overt dependence}. From this figure, an approximately linear temperature dependence of $z$ is observed from around the high-temperature vicinity of the transition toward lower temperatures. Together with the behavior of the correlation length as a function of time at each temperature, we set the upper bound of the temperature range used in the scaling analysis to $T=0.485$ for both $\chi_{\mathrm{SG}}$ and $\chi_{\mathrm{CG}}$.

We then determine the lower bound of the temperature range by the following procedure. We first set a candidate lower bound $T_{\min}$ for the dataset and perform the GPR-based parameter optimization under the constraint that the estimated transition temperature satisfies $T_{\min} \ge T_{\mathrm{c}}$. If the scaling fit fails (e.g., no satisfactory collapse is obtained, or the optimized $T_{\mathrm{c}}$ hits the lower-bound constraint, $T_{\mathrm{c}}=T_{\min}$), we increase $T_{\min}$ and repeat the same optimization. We finally adopt the smallest $T_{\min}$ for which the scaling fit succeeds stably. As a result, the final temperature ranges are $0.440 \le T \le 0.485$ for the SG transition and $0.455 \le T \le 0.485$ for the CG transition.

\begin{figure}[tbp]
    \centering
    \begin{minipage}{0.82\linewidth}
        \centering
        \includegraphics[width=\linewidth]{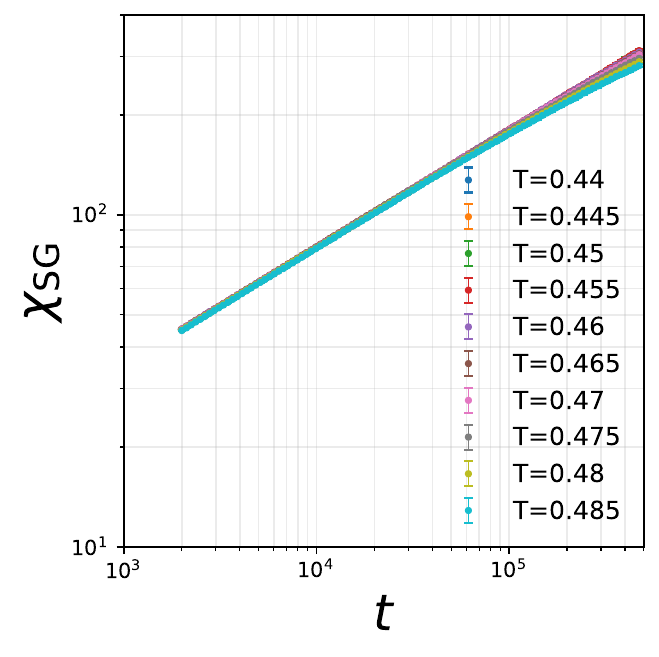}
        \\(a)
    \end{minipage}
    \vspace{2mm}

    \begin{minipage}{0.82\linewidth}
        \centering
        \includegraphics[width=\linewidth]{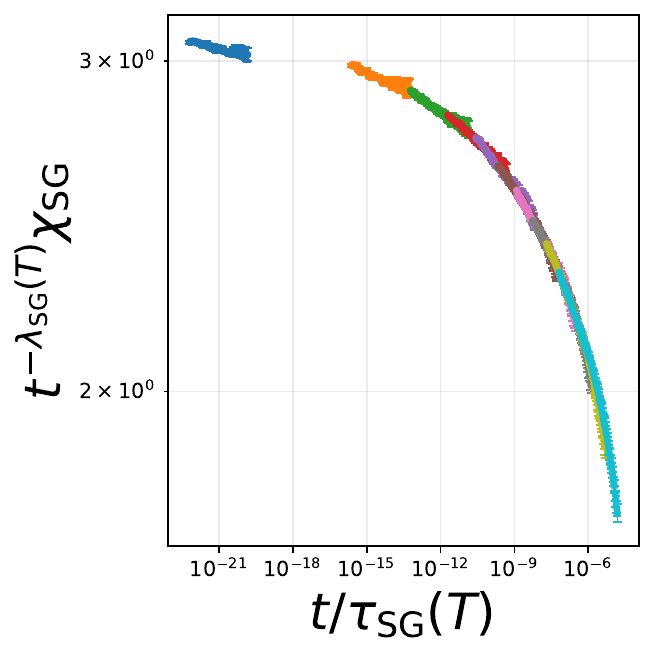}
        \\(b)
    \end{minipage}

    \caption{The results of the dynamical scaling analysis for the 3D $\pm J$ XY model based on Eq.~(\ref{dynamical scaling law for critical temperature SG}): (a) relaxation data for $\chi_{\mathrm{SG}}$ and (b) the scaling plot. The scaling plot in (b) was obtained using the parameters $T_{\mathrm{SG}} = 0.4388(3)$, $\gamma/z\nu = 0.3528(1)$, and $z\nu = 8.67(5)$.}
    \label{3d XY SG scaling result}
\end{figure}

\begin{figure}[tbp]
    \centering
    \begin{minipage}{0.82\linewidth}
        \centering
        \includegraphics[width=\linewidth]{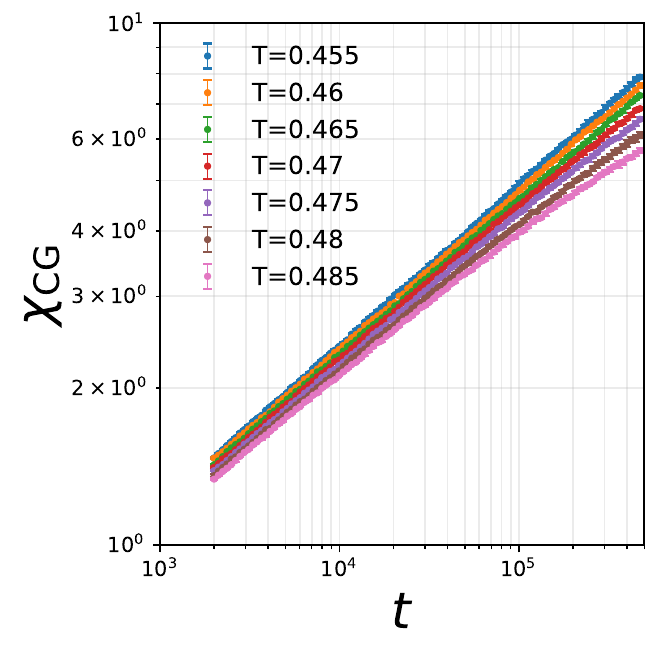}
        \\(a)
    \end{minipage}
    \vspace{2mm}

    \begin{minipage}{0.82\linewidth}
        \centering
        \includegraphics[width=\linewidth]{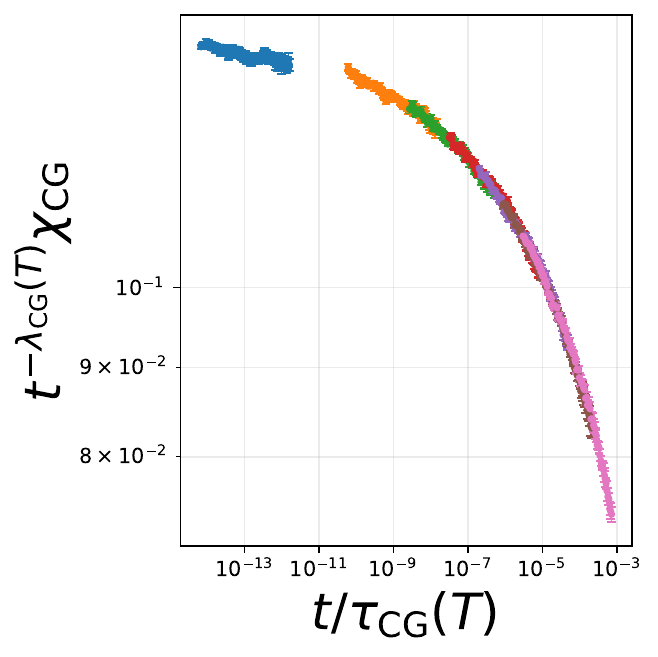}
        \\(b)
    \end{minipage}

    \caption{The results of the dynamical scaling analysis for the 3D $\pm J$ XY model based on Eq.~(\ref{dynamical scaling law for critical temperature CG}): (a) relaxation data for $\chi_{\mathrm{CG}}$ and (b) the scaling plot. The scaling plot in (b) was generated using the parameters $T_{\mathrm{CG}} = 0.4533(7)$, $\gamma/z\nu = 0.3106(4)$, and $z\nu = 6.29(10)$. In this scaling analysis, $z$ is treated as described in Eq.~(\ref{z assumption CG}).}
    \label{3d XY CG scaling result}
\end{figure}

The corresponding scaling results for SG and CG are shown in Figs.~\ref{3d XY SG scaling result} and \ref{3d XY CG scaling result}, respectively. From these analyses, the transition temperatures and the scaling parameters, namely the product of critical exponents $\gamma/z\nu$ and $z\nu$, are estimated. Specifically, for $\chi_{\mathrm{SG}}$ we obtain $T_{\mathrm{SG}}=0.4388(3)$, $\gamma/z\nu=0.3528(1)$, and $z\nu=8.67(5)$; for $\chi_{\mathrm{CG}}$ we obtain $T_{\mathrm{CG}}=0.4533(7)$, $\gamma/z\nu=0.3106(4)$, and $z\nu=6.29(10)$.

Here, $\gamma/z\nu$ can also be estimated independently from the long-time slope of the susceptibility at the transition temperature. Weighted fits to the data in Fig.~\ref{fig:size dependence of 3D XY SG} give $\gamma/z\nu=0.3525(14)$ for the SG susceptibility for $t\geq5\times10^{4}$ and $\gamma/z\nu=0.3109(7)$ for the CG susceptibility for $t\geq2\times10^{4}$. These estimates agree within their error bars with the corresponding dynamical-scaling results, $0.3528(1)$ and $0.3106(4)$, respectively. This agreement in the long-time regime provides an independent check of the validity of the dynamical scaling analysis.

The obtained transition temperatures appeared to show good accuracy compared with previously reported values. The estimates satisfy $T_{\mathrm{SG}}<T_{\mathrm{CG}}$ and are therefore consistent with the spin chirality decoupling picture in SG systems \cite{Kawamura1992}.\par

Although the authors of some previous studies in which the NER method was used did not provide evidence supporting this theory \cite{Yamamoto2004,Nakamura2006}, our analysis of the transition temperature using a similar approach yielded results consistent with it.
In the present work, we consider that the primary origin of the discrepancy from the previous estimates is the choice of the temperature range used in the scaling analysis, namely, restricting the dataset to the range where $z(T)$ exhibits an approximately linear temperature dependence.
To examine this point more closely, we compared our results with those of Ref.~\cite{Nakamura2006} under conditions that mimic the earlier study: we imposed the same lower temperature cutoff ($T\ge 0.460$) and truncated our relaxation data at $1\times10^{5}$~MCS, which was the maximum simulation time used in Ref.~\cite{Nakamura2006}.

\begin{figure}[tbp]
    \centering
    \begin{minipage}{0.82\linewidth}
        \centering
        \includegraphics[width=\linewidth]{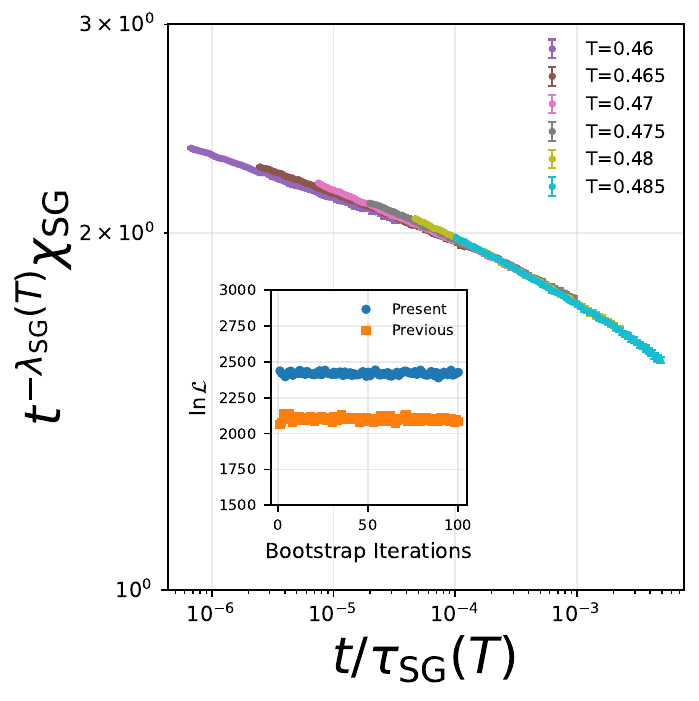}
    \end{minipage}
    \vspace{2mm}

    \caption{Scaling plot of the relaxation data obtained in our study, using the parameters from a previous study~\cite{Nakamura2006}: $T_{\mathrm{SG}} = 0.435$, $\gamma/z\nu = 0.368$, and $z\nu = 6.25$. The inset shows the distribution of the log-likelihood values obtained via bootstrap sampling, comparing the fits using the physical parameters from a previous study~\cite{Nakamura2006} and those obtained in the present study. The clearly separated distributions of the log-likelihood values show that the two scaling behaviors are distinguishable, which supports the physical conclusion drawn in this study.}
    \label{fig:compare nakamura2006}
\end{figure}

Specifically, we fixed the physical parameters $(T_{\mathrm{SG}}, \gamma/z\nu, z\nu)$ either to the values reported in Ref.~\cite{Nakamura2006} or to those obtained in the present study, and for each bootstrap sample we optimized the remaining GPR hyperparameters and compared the resulting log-likelihood values.
The results are shown in Fig.~\ref{fig:compare nakamura2006}, where the main panel presents the dynamical scaling plot obtained from our relaxation data truncated at $1\times10^{5}$~MCS, and the inset shows the distribution of the log-likelihood values over bootstrap samples for the two fixed-parameter fits.

Even under this restriction to $1\times10^{5}$~MCS, the fits using our parameter set yield, on average, higher log-likelihood values than those using the parameters of Ref.~\cite{Nakamura2006}.
These observations support the validity of our analysis strategy, in particular the use of only the temperature range in which the assumed linear behavior of $z(T)$ is satisfied.

\subsection{Estimation of critical exponents}
\label{subsection: Critical Values Estimated by $z$ and Dynamical Scaling}
Using the transition temperatures obtained from the analysis in Sec.\ref{subsection: Critical Temperatures from Dynamical Scaling Analysis}, we evaluate the dynamical exponent $z$ at the critical temperatures. Based on these results and the products of critical exponents obtained in Sec.\ref{subsection: Critical Temperatures from Dynamical Scaling Analysis}, we assess the critical behavior.

\begin{figure}[tbp]
    \centering
    \begin{minipage}{0.82\linewidth}
        \centering
        \includegraphics[width=\linewidth]{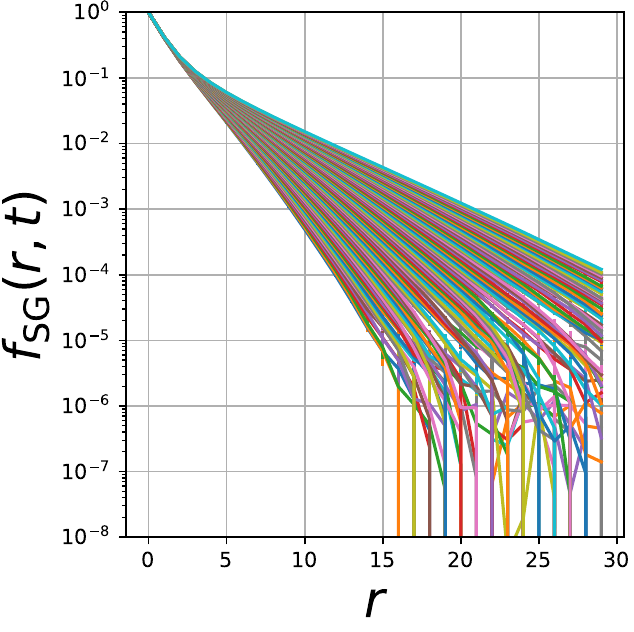}
        \\(a)
    \end{minipage}
    \hfill
    \begin{minipage}{0.82\linewidth}
        \centering
        \includegraphics[width=\linewidth]{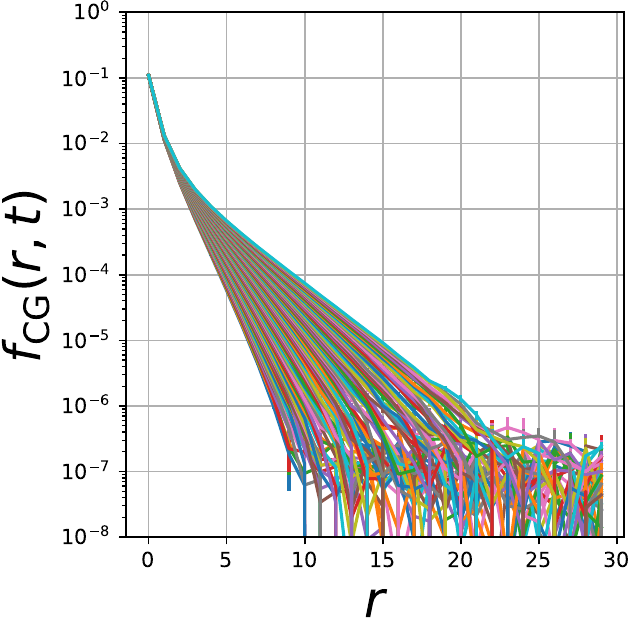}
        \\(b)
    \end{minipage}

    \caption{The dynamical SG and CG correlation functions of the 3D $\pm J$ XY model at the critical temperature for various Monte Carlo steps $t$: (a) analysis of the SG transition and (b) analysis of the CG transition.}
    \label{fig:correlation function 3D XY SG}
\end{figure}

\begin{figure}[tbp]
    \centering
    \begin{minipage}{0.82\linewidth}
        \centering
        \includegraphics[width=\linewidth]{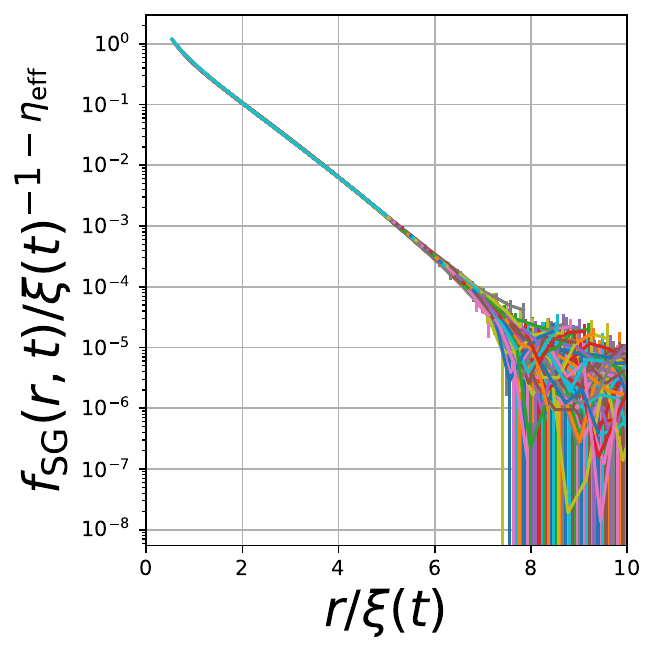}
        \\(a)
    \end{minipage}
    \vspace{2mm}

    \begin{minipage}{0.82\linewidth}
        \centering
        \includegraphics[width=\linewidth]{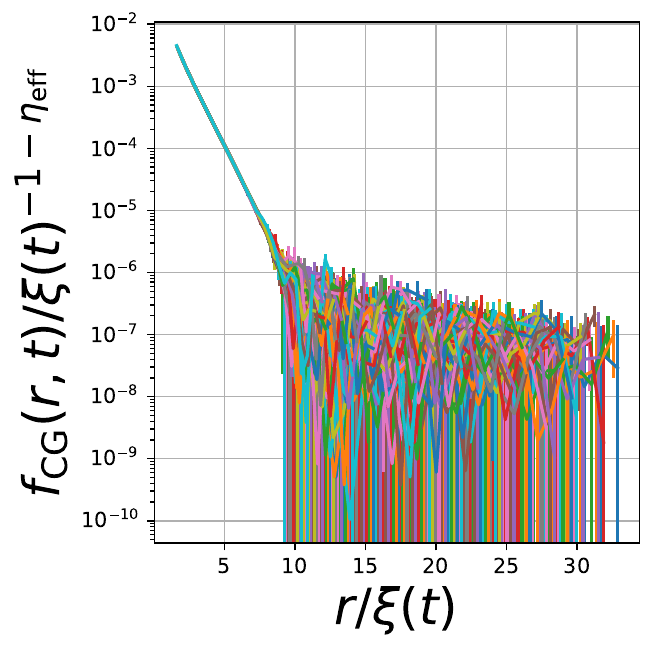}
        \\(b)
    \end{minipage}

    \caption{The scaling plots of the SG and CG correlation functions in the 3D $\pm J$ XY model: (a) analysis of the SG transition with $\eta_{\mathrm{eff}} = 0.307(2)$, and (b) analysis of the CG transition with $\eta_{\mathrm{eff}} = 0.555(2)$.}
    \label{fig:scaling plot correlation function 3D XY SG}
\end{figure}

\begin{figure}[tbp]
    \centering
    \includegraphics[width=0.98\linewidth]{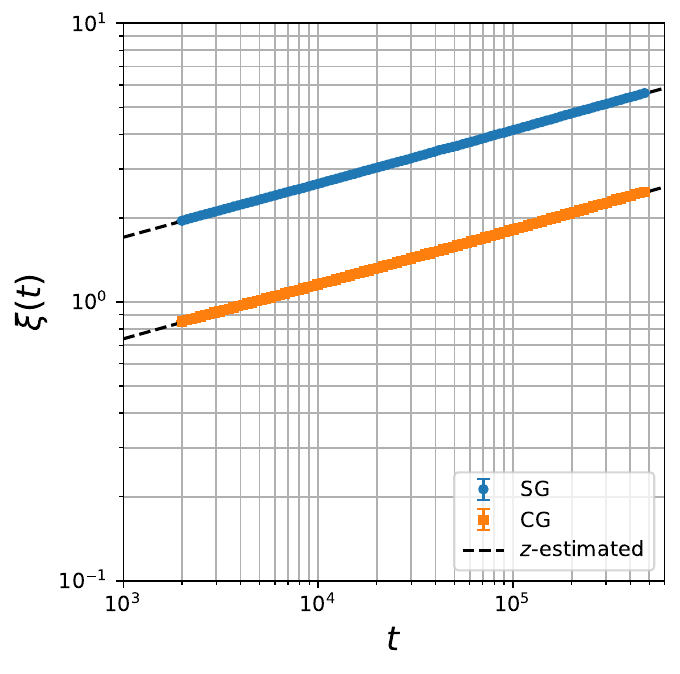}
    \caption{The SG and CG correlation lengths in the 3D $\pm J$ XY model. The dotted line indicates the slope corresponding to the estimated dynamic exponent $z$.}
    \label{fig:correlation length 3D XY SG}
\end{figure}

The results of the correlation function computed at the transition temperatures are shown in Fig. \ref{fig:correlation function 3D XY SG}. We subsequently perform a dynamical scaling analysis based on Eq. (\ref{correlation dynamical scaling law}). In an optimization procedure, we discarded short-range correlations ($r<3$ for $f_{\mathrm{SG}}$ and $r<4$ for $f_{\mathrm{CG}}$) and data of small values ($f_{\mathrm{SG}}(r,t)<10^{-5}$ and $f_{\mathrm{CG}}(r,t)<2\times10^{-6}$).  The initial values of $\{\xi(t)\}$ were set to those of correlations obtained by the second moment method, represented by Eq. (\ref{eq: SG second moment correlation length}) for SG and Eq. (\ref{eq: CG second moment correlation length}) for CG. The scaling plots by inference correlation length are shown in Fig. \ref{fig:scaling plot correlation function 3D XY SG}.
Figure \ref{fig:correlation length 3D XY SG} shows the scaling result based on the dynamical scaling law in Eq. (\ref{correlation dynamical scaling law}).

From the inference results and Eq. (\ref{z definition}), the value of $z$ is evaluated as $z=5.191(3)$ for the SG case and $z=5.098(4)$ for the CG case. The uncertainties were quantified as the standard deviations of the bootstrap distributions derived from 128 resamplings\cite{Efron1979}. These results and the previously mentioned dynamical scaling analysis in Sec. \ref{subsection: Critical Temperatures from Dynamical Scaling Analysis} can be used to estimate the critical exponents. These results are compared with those of previous studies in Tables \ref{tab:3DXY SG Result} and \ref{tab:3DXY CG Result}.\par

The obtained results show improved accuracy for both the transition temperature and the critical exponents compared with previously reported findings.
In addition, the dynamical exponent $z$ was found to be lower in the CG case than in the SG case, which supports the observation that finite-size effects are more pronounced in the CG sector (Fig.~\ref{fig:size dependence of 3D XY SG}).

\begin{table}[tbp]
    \centering
    \caption{Analysis results of the SG transition in the 3D $\pm J$ XY model compared with the corresponding results reported in previous studies.}
    \begin{tabular}{ccccc}\hline
    Reference & $T_{\mathrm{SG}}$ & $\gamma$ & $z$ & $\nu$\\\hline
    \cite{Granto2004} & 0.39(2) &  & 4.4(3) & 1.2(2) \\
    \cite{Yamamoto2004} & 0.455(15) & 1.70(14) & 5.6(6) & 0.85(8) \\
    \cite{Nakamura2006} & 0.435(15) & 2.30(25) & 5.0(1) & 1.25(15) \\
    Present work & 0.4388(3) & 3.05(2) & 5.191(3) & 1.67(1) \\\hline
    \end{tabular}
    \label{tab:3DXY SG Result}
\end{table}

\begin{table}[tbp]
    \centering
    \caption{Analysis results of the CG transition in the 3D $\pm J$ XY model compared with those in previous studies.}
    \begin{tabular}{ccccc}\hline
    Reference & $T_{\mathrm{CG}}$ & $\gamma$ & $z$ & $\nu$ \\\hline
    \cite{Kawamura2001} & 0.39(3) &  & 7.4(10) & 1.2(2) \\
    \cite{Yamamoto2004} & 0.467(10) & 1.41(8) & 6.3(5) & 0.74(4) \\
    \cite{Nakamura2006}& 0.445(15) & 1.65(2) & 5.1(2) & 1.05(10) \\
    Present work & 0.4533(7) & 1.95(3) & 5.098(4) & 1.23(2) \\\hline
    \end{tabular}
    \label{tab:3DXY CG Result}
\end{table}

\section{Summary and Discussion}
\label{sec: summary and discussion}
We primarily aimed to validate a method for calculating the dynamical exponent from the correlation function, thereby enabling an analysis of SG systems. The obtained dynamical exponent appeared to be reliable, showed good accuracy, and was not limited to SG systems. The results confirm the reliability of this method. Using this approach, we analyzed the 3D $\pm J$ XY model. Compared with conventional methods, our method achieved a higher precision in evaluating the dynamical exponent. Through this analysis, we also examined the temperature dependence of $z(T)$ near both the SG and CG transitions. In addition, the estimated transition temperatures are consistent with the spin chirality decoupling picture in SG systems.

Although some previous studies using the NER method yielded nearly coincident estimates of the SG and CG transition temperatures \cite{Yamamoto2004,Nakamura2006}, our analysis using a similar method yielded estimates satisfying $T_{\mathrm{SG}}<T_{\mathrm{CG}}$. However, the estimates obtained from the SG $\chi$--$t$ scaling are sensitive to the choice of the temperature range used in the analysis. In the present work, a satisfactory scaling collapse is obtained by restricting the analysis to the temperature range in which the assumed relation $z(T)\propto 1/T$ is numerically observed. The resulting estimates should therefore be understood within the adopted scaling form and temperature range.

To examine the $\chi$--$t$ scaling results from an independent perspective, we also performed $\chi$--$\xi$ scaling for both SG and CG, following the approach of Refs.~\cite{Nakamura2016,Nakamura2019}. For CG, two scaling branches corresponding to the high- and low-temperature sides were obtained. For SG, however, systematic offsets remained among the temperature series, and a satisfactory collapse onto a single scaling function was not obtained. Thus, $\chi$--$\xi$ scaling did not provide an independent validation of the SG results.

The contrast between the SG and CG scaling results, together with the sensitivity of the SG $\chi$--$t$ scaling to the analyzed temperature range, suggests that the physics in the intermediate temperature region $T_{\mathrm{SG}}<T<T_{\mathrm{CG}}$ may affect the SG relaxation and contribute to the instability of the scaling analysis. The specific mechanism responsible for this behavior is not clear from the present results, and the relation cannot be established by the present analysis alone. Clarifying the relation between the intermediate temperature region and the instability of the SG scaling analysis remains an important subject for future work.

The correlation lengths estimated using our method are affected by finite-time corrections. Because we do not extrapolate the scaling regime to $t\rightarrow\infty$ when calculating the dynamical exponent, the proposed method can be difficult to apply to cases with strong fluctuations. An example of such a situation is the analysis of SG transitions near multi-critical points in the 3D $\pm J$ Ising model\cite{Terasawa2023}. Making the method more widely applicable will require the development of an extrapolation method for $t\rightarrow\infty$. A reliable and systematic extrapolation scheme has recently been proposed in the NER method\cite{Osada2024}. We expect that using such a scheme will lead to solutions to such problems. This work is also a future task.

\section*{Acknowledgments}
 This work was supported by JST SPRING, Grant Number JPMJSP2131. The authors are also grateful to the Supercomputer Center at the Institute for Solid State Physics, University of Tokyo, for use of their facilities.

\makeatletter
\let\auto@bib@innerbib\@empty
\makeatother

\makeatletter
\global\let\auto@bib\@empty
\global\let\write@bibliographystyle\relax
\makeatother

\end{document}